\documentstyle[11pt,newpasp,twoside,epsf]{article}
\def\edcomment#1{\iffalse\marginpar{\raggedright\sl#1\/}\else\relax\fi}
\reversemarginpar

\newcommand{\teff}{$T_{\rm eff}$}

\newcommand{\lsun}{{\rm L}_{\mbox{\rm \sun}}}

\newcommand{\ha}{H$\alpha$}

\newcommand{\HeI}{He\,{\sc i}}
\newcommand{\HeII}{He\,{\sc ii}}

\newcommand{\FeII}{Fe\,{\sc ii}}

\newcommand{\OI}{O\,{\sc i}}

\newcommand{\OIII}{O\,{\sc iii}}

\newcommand{\SiIV}{Si\,{\sc iv}}

\newcommand{\CIV}{C\,{\sc iv}}

\newcommand{\NV}{N\,{\sc v}}

\newcommand{\ergs}{erg\,s$^{-1}$}
\newcommand{\ccm}{cm$^{-3}$}
\newcommand{\kms}{km\,s$^{-1}$}

\newcommand{\be}{B[e]}
\begin{document}
\title{The Connection with B[e] stars}
 \author{Franz-Josef Zickgraf}
\affil{Hamburger Sternwarte, Gojenbergsweg 112, 21029 Hamburg, Germany}

\begin{abstract}
The characteristics of the various types of B[e] stars are discussed 
and compared with those of classical Be stars. Both groups of stars 
are characterized by the presence of emission lines
in their spectra, in particular of hydrogen.  However, 
there are also significant differences between these classes. 
Classical Be stars lack hot circumstellar dust and 
strong forbidden low-excitation emission lines, which are typical
characteristics produced by B[e]-type stars.
While classical Be stars are a rather uniform group of early-type stars, 
B[e]-type stars form  a quite heterogeneous group, very often 
of poorly known evolutionary status, comprising such diverse types of 
objects as near main-sequence objects, evolved low-mass proto-planetray nebulae and 
massive evolved hot supergiants. Even  pre-main sequence Herbig
Ae/Be stars sometimes find their way into the group of B[e] stars. 
However, despite these dissimilarities 
classical Be stars and B[e]-type stars, share a 
common property, namely the non-sphericity of their circumstellar envelopes. 
\end{abstract}


\section{Introduction}
The existence of classical Be stars has been recogized 
for more than a century. The history of \be\
stars on the other hand is much shorter. About 30 years ago
Geisel (1970) first reported a correlation between 
stars for which the continua exhibit
excess radiation at infrared wavelengths $> 5\mu$m and stars having spectra 
which exhibit low-excitation emission lines. This marked the beginning of the
investigation of the enigmatic class of B[e] stars although then these
stars were not yet know under this acronym.
At about the same time Wackerling (1970) and  Ciatti et al. (1974) also
realized the existence of a group of hot emission line objects with 
``abnormal'' spectra and forbidden emission lines for which they introduced 
the designation 
BQ[~] stars. Allen \& Swings (1972) and in particular 
Allen \& Swings (1976) systematically investigated  this new class of stars
which they described as peculiar hot emission line stars with infrared excesses.
They found that these stars ``... form a group spectroscopically discerned 
from normal Be stars and planetary nebulae...'' which contains ``objects 
ranging from almost conventional Be stars to high density planetary nebulae''. 
The dominant spectral features were found to be emission lines of singly 
ionized iron. 
The final step towards the \be\ stars was made by 
P. Conti during the general discussion in the 1976 IAU symposium entitled
 ``Be and Shell Stars.'' He suggested that 
   `` ... A second class of objects would be those B-type stars which show 
   forbidden emission lines and I would suggest that we classify these as B with 
   a small e in  brackets B[e], following the notation for forbidden lines''
(Conti 1976).

In the following sections I will first describe in more detail the defining 
characteristics of what nowadays we call ``\be '' stars, and then consider the 
various object classes which constitute the inhomogeneous group of 
\be\ stars. Finally, I will discuss
in some detail the connection between \be\ stars and classical Be stars.

\section{Defining characteristics of B[e] stars}
Extensive infrared surveys of galactic early-type emission line stars, 
e.g. by Allen (1973), (1974), and Allen \& Glass (1975), revealed that
two populations of emission line stars exist:
(a) emission-line stars with normal stellar infrared colours
comprising the classical Be stars and other types of objects like
normal supergiants, Luminous Blue Variables (LBVs), and S-type symbiotics;
and (b) ``peculiar'' emission-line stars with IR excesses due to 
hot circumstellar dust.

Following Conti's suggestion  the latter  group of
peculiar and dusty Be stars is now usually called
B[e] stars, i.e. they are early-type emission lines stars with low-excitation 
lines, forbidden lines, and  hot dust visible in the near and mid infrared.

As we will see in the next section a more precise statement is that these 
stars show the  ``B[e] phenomenon'' which is characterized by spectroscopic
properties and by the continuum energy distribution as follows:

\begin{itemize}
   \item [1.]Spectroscopic characteristics:
     \begin{itemize}
       \item [1.1]strong Balmer emission lines;
       \item [1.2]low-excitation permitted emission lines,
        predominantly of singly ionized metals, e.g. \FeII;
       \item [1.3] forbidden emission lines of [\FeII ] and [\OI ];
       \item[1.4] higher ionization emission lines can be present, e.g. 
       [\OIII ], \HeII ;
     \end{itemize}  
    \item [2.]continuum energy distribution:
      \begin{itemize}       
        \item [2.1] Continuum distribution of an early-type star in the 
	            visual(/UV);
        \item [2.2] strong near/mid infrared excess due to 
              hot circumstellar dust with  temperatures around 
	      $T \sim 500-1000$\,K.
      \end{itemize}  
\end{itemize}  

\begin{figure}[tbh]
\plotone{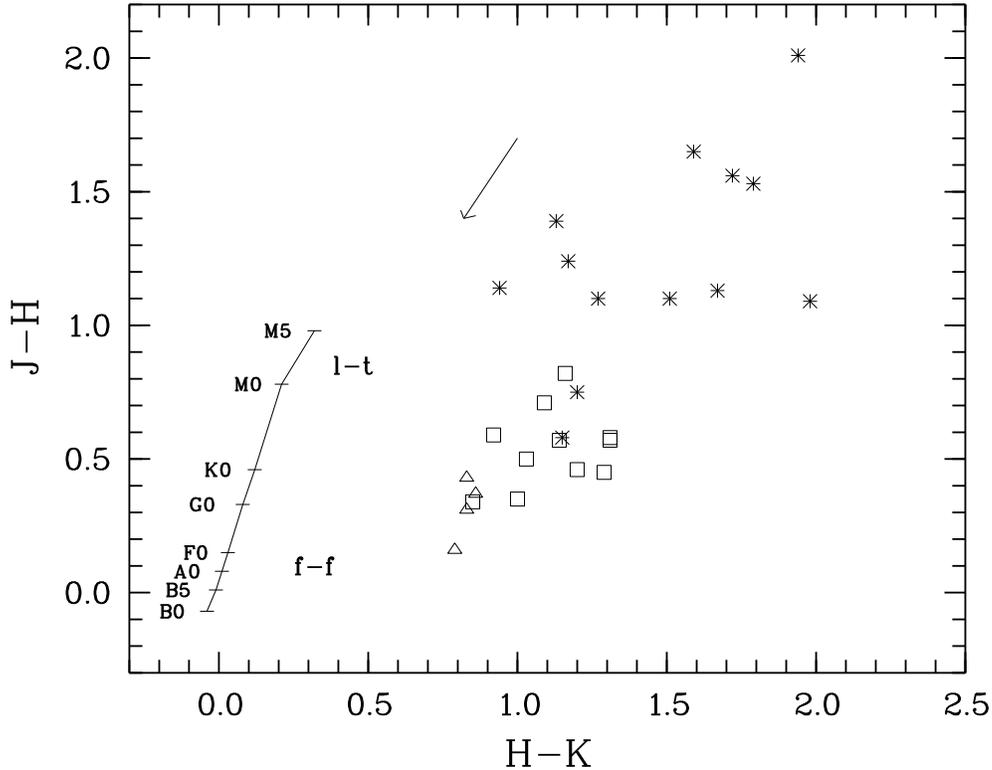}
\caption[]{$(J-H)-(H-K)$ diagram for B[e]-type stars. 
The different symbols denote B[e] stars in the Milky Way (asterisks), 
in the LMC (squares) and in the  SMC (triangles).
The location
of stars  with normal colours is indicated by the solid line. ``f-f'' denotes
the approximate location of stars with IR excess due to $f-f$ emission from
stellar winds. ``l-t'' marks the area occupied by objects with late-type
companions. The arrow gives the reddening line for $A_V = 3$.
}
\label{jhk}
\end{figure}
        
Stars showing the \be\ phenomenon form a  distinct group of objects in 
$(J-H)-(H-K)$ and $(H-K)-(K-L)$ diagrams (cf. Fig. 1). One can therefore state that
in connection with forbidden emission lines hot dust is {\it the} defining 
characteristic of B[e] stars. Hence, a star with forbidden emission lines
but without thermal emission due to hot circumstellar dust
(such as LBVs in certain phases), should not be classified as 
\be\ star.  
With respect to classical Be stars the presence of circumstellar dust is  
a  distinctive feature because these stars definitely lack this property.  

Linear polarization measurements showed that \be\ stars of
all sub-types defined in the next section are characterized additionally by 
a non-spherical distribution of circumstellar
scattering particles. Intrinsic polarization was observed e.g. by Coyne \& Vrba
(1976),
Barbier \& Swings (1982), Zickgraf \& Schulte-Ladbeck (1989), and  
Schulte-Ladbeck et al. (1992).
Oudmaijer \& Drew (1999) obtained spectropolarimetry of B[e] and Herbig Be 
stars. They detected non-sphericity by polarization changes 
across the H$\alpha$ emission line in all B[e]-type stars of their sample.

In classical Be stars polarization is due to  Thomson scattering in the
ionized circumstellar disk. Two scattering mechanisms are responsible for the 
observed polarization of \be\ stars: scattering by dust particles and 
Thomson scattering anaolgous to classical Be stars. 
Both mechanisms may contribute simultaneously in \be\ type stars (cf. 
Zickgraf \& Schulte-Ladbeck 1989)	 

As discussed by Lamers et al. (1998) the  characteristics of the B[e] 
phenomenon can also be formulated in terms of physical conditions:
\begin{itemize}
\item[(a)] The strong Balmer emission lines imply very large emission measures
($EM$) of the singly ionized gas above the stellar continuum forming region.
Typically, for a supergiant \be\ star (see below) with \ha\ luminosities of 
$10^{37}$ to $10^{38}$\,\ergs, the $EM$s are on the order of
$10^{62}$ to  $10^{63}$  \ccm . For less luminous stars, such as pre-main sequence B
stars showing the B[e] phenomenon, the emission measure is about $10^{57}$  \ccm .
\item[(b)] The presence of emission lines of low-ionization metals like \FeII\
indicates a temperature of $\sim$ 10$^4$\,K in the emitting region. 
\item[(c)] The presence of forbidden emission lines of low excitation metals
such as [\FeII ] and [\OI ] indicates a geometrically extended envelope so that
there is a large amount of low density gas. Applying the diagnostics described 
by Viotti (1976) leads to densities of the [\FeII ] emitting region of $N_e <
10^{11}$\,\ccm .  
\item[(d)] According to Bjorkman (1998), 
infrared excesses from cool dust ($T_{rm d}$ = $\sim$500-1000\,K)
indicate a circumstellar density 
of $\rho \ge 10^{-18}$\,gm\,\ccm\ at distances where the dust
temperature can equilibriate ($\ge 500$ to 1000\,R$_{\ast}$). 
\end{itemize}

Taking into account that the circumstellar environments of \be\ stars are 
non-spherical, these conditions are consistent with circumstellar densities on
the order of 10$^{9}$ to 10$^{10}$\,\ccm . This is supported also by
2.3\,$\mu$m CO overtone emission observations by MacGregor et al. (1988a,b) 
who derived similar densities for the molecular 
emission regions. The existence of dust is most likely related to the existence
of molecules. The formation of molecules in turn drives the condensation of
dust particles. Both types of matter require high densities. In contrast,
the envelopes of Be stars may have densities
too low to allow condensation of dust particles.

\section{B[e] stars: a stellar melange}
The definition of B[e] stars as given in the previous section
describes certain physical conditions in terms of excitation and density 
in the circumstellar environment rather than intrinsic object properties.
Because similar circumstellar conditions can prevail in the surroundings 
of objects belonging to intrinsically different classes it is not 
surprising that  ``B[e]'' stars do  {\bf not} form a homogenous group of 
objects,  but rather a melange of various classes. 

This was already realized by Allen \& Swings (1976) who distinguished 
three groups of peculiar Be stars with infrared excesses, namely
     (a) group 1: few emission lines, not always forbidden lines, almost conventional
      Be stars,
     (b) group 2: ``most distinctive group'', spectra with permitted and forbidden
      \FeII\ emission, and  
    (c)  group 3: additionally emission lines of higher ionization stages 
     ($IP > 25$eV).  
The phenomenological grouping indicated that \be\ stars are not a unique class 
of objects but comprise members of different classes which share the 
common property of showing the \be\ phenomenon  (for a detailed discussion 
of the various object classes cf. Zickgraf 1998).

Actually, it would be desirable to determine the intrinsic stellar parameters 
in order to find the position of the \be\ stars in the H-R diagram. This would 
permit one to constrain the likely evolutionary status and therefore determine
the objects' B[e] classification types. However, in practice this
turns out to be difficult or even impossible. 
In many cases photospheric absorption features are
absent or at most weakly discernible. It is therefore difficult to determine 
reliable effective temperatures. In many cases only the stellar continuum energy
distribution allows to estimate the star's \teff , yielding rather uncertain 
results.
Interstellar reddening increases the problem.  Likewise, unknown distances 
lead to uncertain  luminosities. Therefore  the  evolutionary status of many 
B[e] stars is unknown.
Zorec (1998) collected  distances and luminosities for galactic B[e]-type 
stars in order to establish the H-R diagram  for galactic B[e] stars.
But even with known $T_{\rm eff}$ and $L$ values, 
\be\ stars often offer obstacles to a determination of 
their evolutionary status, as I will discuss below for some
near-main sequence objects. 

In some cases more or less reliable information about the evolutionary status 
can be obtained.  Some stars were found  to represent objects in a post-main 
sequence phase of the evolution of massive stars. Others are obviously 
intermediate mass pre-main sequence Herbig Ae/Be stars, while still others 
are in late stages of the evolution of low-mass stars.

As an important result of the Workshop on B[e] stars held in  Paris, 1997,
Lamers et al. (1998)  discussed an improved classification of B[e] stars and
suggested several different subclasses of B[e] stars:
 \begin{itemize}
     \item[a)]  evolved high-mass stars with $L \ge 10^4\lsun$: B[e] supergiants
       $\rightarrow$ sgB[e]
     \item [b)] intermediate mass pre-main sequence stars: Herbig Ae/Be stars
      $\rightarrow$ HAeB[e]\\
     in particular: {\it isolated} HAeB[e]
     \item[c)]  evolved low-mass stars: \\
     Compact low-excitation proto-planetary nebulae $\rightarrow$ cPNB[e]
     \item [d)] D-type symbiotic stars $\rightarrow$ symB[e]
     \item[e)] ``unclassified'' B[e] stars $\rightarrow$ unclB[e]
  \end{itemize}
As a further possible class one can add a group of main-sequence or near-main 
sequence stars, MS\be , which could represent a link with classical Be stars.

Lamers et al. stressed that a unique classification is not always 
possible because the assignment to a class is typically ambiguous. It is therefore
not surprising that their group of stars of type unclB[e] is the largest. 

\section{The connection: B[e] vs. Be}
In this section I will discuss in more detail the connection between Be stars
and the two subclasses of low-luminosity B[e] stars and high-luminosity 
B[e] supergiants. The dividing line between both classes is set around $L
\simeq 10^4\,\lsun$.

\subsection{Low-luminosity/(near)-MS  B[e] stars}
In the H-R diagram constructed by 
Zorec (1998) several B[e] stars are found close to or on the main sequence.
These stars are of particular interest to resolving the question of their 
connection 
with classical Be stars. Among these objects are e.g.  MWC\,84, and HD\,51585, 
which were classified as cPPNB[e] by Lamers et al. (1998), and HD\,163296, 
HD\,31648, and HD\,190073 which are probably HAeB[e]-type stars. They
belong to classes which are not related to classical Be stars.

This is different for the two \be\ stars HD\,45677 (FS CMa) and 
HD\,50138. In the H-R diagram  they are located in the same region as
classical Be stars, and they exhibit spectroscopic similarities with
this class. For these two near-main sequence objects  parallaxes are known 
from HIPPARCOS (cf. Zorec 1998) and therefore their distance and luminosity 
are well known.  

\subsubsection{HD\,45677}
This is one of the best-studied galactic B[e] stars. 
Allen \& Swings (1976) describe 
HD\,45677 as a kind of ``proto-type'' of their group 2 (s. above) 
and hence it is often regarded as a ''proto-type'' B[e] star.
Its spectral type is B2(III-V)e. The spectrum was described in detail e.g. by 
Swings (1973), de Winter \& van den Ancker (1997), and  by 
Andrillat et al. (1997). A detailed  NLTE analysis by Israelian et al. 
(1996) yielded the stellar parameters $\log g = 3.9$ and $T_{\rm eff} = 
22\,000$\,K. 

Swings (1973) discussed the emission line spectrum of HD\,45677. He found 
double-peaked \FeII\ emission lines  with a line splitting of $\Delta 
v = 32$\,\kms , and single-peaked forbidden [\FeII ] lines. 
He interpreted the observations in terms of a rotating equatorial disk.
The non-spherical environment is confirmed by visual and UV polarization data,
which indicate that the star is viewed edge-on through a 
dusty disk (Coyne \& Vrba 1976, Schulte-Ladbeck et al. 1992). 
The existence of a disk should be correlated with a high stellar rotational 
velocity for which Swings \& Allen (1971) had estimated in fact  $v \sin i 
\approx 200$\,\kms\  which would be comparable with velocities measured 
for classical Be stars. The NLTE analysis of 
Israelian et al. (1996), however, yielded only $v \sin i  \approx  70$\,\kms .
It remains therefore unclear whether rapid rotation is responsible for the 
formation of the disk.

Although the distance of HD\,45677 is known its evolutionary status is still
under discussion. The location in the H-R diagram suggests that it is near the
main sequence. In this respect it could be considered as extreme Be star. 
However, Grady et al. (1993) 
detected  mass accretion in IUE spectra which indicates that it is still in a
pre-main sequence phase of evolution, i.e. a  Herbig Be star of the type
HAeB[e]. This classification is somewhat doubtful because  no association 
with a nebula is present. de Winter \& van den Ancker (1997) therefore 
interpreted the observations in view of the isolated position in the sky in 
terms of a young 
object, but not in the sense of being pre-main sequence. It should be noted,
however, that the isolated location does not necessarily contradict a 
Herbig Be classification. Grinin et al. (1989, 1991) discuss the existence 
of isolated Herbig stars, of which HD\,45677 could be a member.

\subsubsection{HD\,50138}
HD\,50138 was considered by Allen \& Swings (1976) as a  group 1 object 
and  they described this star as a kind of extreme Be star. For recent studies
cf. Pogodin (1997) and Jaschek \& Andrillat (1998). 
Its spectral type is B6III-IV, and in the H-R diagram it is located 
close to the main sequence. Houziaux (1960) determined a rotational velocity 
of $v \sin i \approx 160$\,\kms . Like HD\,45677 it is not 
associated  with a nebula, but exhibits spectral and polarimetric 
characteristics similar to young stellar objects. It could therefore also 
be a HAeB[e]-type star. 

The conclusion for the (near)-main sequence B[e] stars is that their
 $T_{\rm eff}$ and $L$ are comparatively well known, 
and that in some respects they are
similar to Be stars. Like the latter group, they posses
disk-like circumstellar structures and may also rotate rapidly -- although this
question is not yet settled. 
However, their evolutionary status is still a controversial issue. 
Despite their well known distances and effective 
temperaturesm it is still not clear whether they are a kind of extreme Be stars 
or, alternatively, 
pre-main sequence HAeB[e] stars. The problem is that near the location of
these stars in the H-R diagram the birthline of Herbig stars 
reaches the main sequence (Palla \& Staller 1993) and hence a separation of 
true main sequence from pre-main sequence objects is difficult.

\subsection{ B[e] supergiants vs. Be stars}
As discussed in the previous section, galactic B[e] stars are a
mixture of different classes. 
Only for a few of these B[e] stars luminosities are known to be as high as those 
of supergiants, like e.g. 
CPD-52$^{\circ}$9243 (Swings 1981, Winkler \& Wolf 1989),
MWC 300 (Wolf \& Stahl 1985),
MWC 349A (Cohen et al. 1985),
GG Car (McGregor et al. 1988a, Lopes et al. 1992),
HD 87643 (Oudmaijer et al. 1998), and
MWC 137 (S 266) (Esteban \& Fernandez et al. 1998).

However, despite many efforts the classification remains often uncertain,  
and confusion with other classes is still an issue,
e.g. as for the
stars MWC 300, HD 87643, MWC 349A, and MWC 137. These stars 
have alternatively been
classified as HAeB[e]-type stars with lower luminosities than supergiants 
(cf. references in Lamers et al. 1998).  

For extragalactic \be\ stars the situation is different. Until now the only
galaxies in which \be\ have been found (except our Milky Way) are the 
Magellanic Clouds (MCs). Presently 15 B[e] supergiants known in the MCs.
For a list of these stars cf. Lamers et al. (1998). 
Compared to the Milky Way the advantage of the MCs is that their 
distances are well known and hence the luminosities of the \be\ stars are also 
known. The location of the \be\ stars in the H-R diagram  indicates that they 
are evolved massive post-main sequence objects.
Most  of the sgB[e] in the MCs have  
luminosities on the order of $L \approx 10^5 - 10^6$ L$_{\odot}$ 
(e.g. Zickgraf et al. 1985, 1986, 1989).
Recently, the luminosity range was extended 
downward to $L \approx 10^4$L$_{\odot}$ 
(Zickgraf et al. 1992, Gummersbach et al. 1995) suggesting that a
transition to lower-luminosity near-main sequence objects could exist.

\begin{figure}[tbh]
\plotone{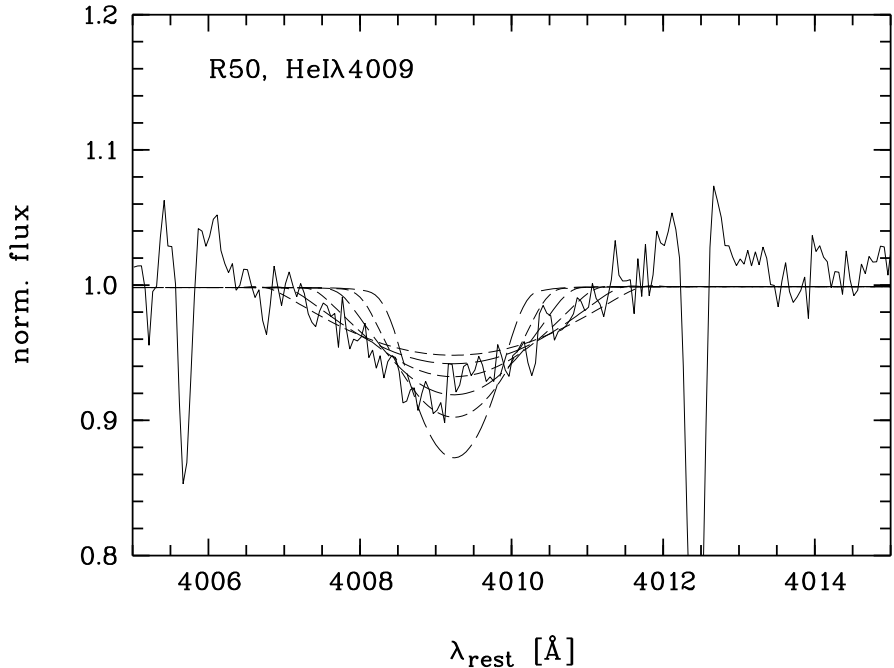}
\caption{Section of the spectrum of R\,50 in the SMC (spectral resolution $\sim
20\,000$). Overplotted are rotationally broadened synthetic line profiles of
\HeI$\lambda$4009\AA\ for rotational velocities between 75\,\kms\ and 200\,\kms\
in steps of 25\,\kms . The best fit yields $v\sin i \simeq 150$\,\kms .  }
\label{rot}
\end{figure}

Spectroscopically \be\ supergiants are characterized by hybrid spectra. This term 
describes the simultaneous presence of broad (1000-2000\,\kms ) high excitation absorption 
features of \NV , \CIV , and \SiIV\  in the satellite UV, or of \HeI\ in 
the visual wavelength region indicative of a 
hot high velocity wind component. At the same time 
narrow ($\la 100$\,\kms ) \be -type low-excitation emission-lines  of \FeII , 
[\FeII ], and [\OI ] are observed, 
a fact which suggests a cool, slow wind component.
Likewise, molecular emission bands of TiO in the visual and 
CO overtone bands in the near infrared are observed (Zickgraf et al. 1989,
McGregor et al. 1988b). These observations indicate the presence of two 
wind regions with basically different physical conditions.
	
The observed properties were explained by an empirical model by Zickgraf et 
al. (1985). These authors
invoked a two-component stellar wind with a bipolar wind structure.
In the polar region a fast radiation-driven CAK-type wind, similar to those
of normal OB supergiants, prevails. This component exhibits
velocities $\sim 1000-2000$\,\kms .
In the equatorial region region a slow, cool, and dense wind is present with 
outflow velocites on the order of $\sim 100$\,\kms . The equatorial 
outflow is concentrated in a disk-like configuration, similar to models 
for classical Be stars. However, in classical Be stars the polar wind is 
significantly less pronounced due to these stars' lower luminosities. Likewise, 
supergiants are closer to the Eddington limit. They have values of 
$\Gamma = 1-(g_{\rm eff}/g_{\rm grav}) \ge 0.3-0.5$. Therefore even moderate rotation of 100-200\,\kms\ is
near the break-up velocity (Zickgraf et al. 1996). In fact 
there is an indication for fast rotation 
for at least one case  of a \be\ supergiant, namely R\,50 in the SMC. This star rotates with a velocity on the order of 
$v\sin i \simeq$ 150\,\kms\ (cf. Fig. 2) and thus at $\ga 60$\% of its break-up velocity.

Linear polarization measurements support the assumption of non-spherical 
circumstellar environments for \be\ supergiants. Magalhaes (1992) and 
Schulte-Ladbeck et al. (1993) detected intrinsic polarization in several 
B[e] supergiants.  Schulte-Ladbeck \& Clayton (1993) obtained
spectropolarimetry of Hen S22 in the LMC and detected intrinsic polarization 
due to electron scattering in a circumstellar disk.

\section{Summary}
The group of B[e] stars is an  inhomogeneous class of objects
comprising intrinsically different classes characterized by 
permitted and forbidden low-excitation line emission and hot circumstellar dust. 
In contrast, classical Be stars are more or less normal B stars with photospheric 
absorption lines and superimposed Balmer emission lines, plus occasionally 
\FeII\ emission lines. B[e] stars in many cases show no or only weak 
photospheric absorption lines. Forbidden lines on the other hand are 
not typical for classical Be stars. Likewise, circumstellar dust signatures 
are not found in  classical Be stars but are a defining characteristic of \be\
stars. High stellar rotational velocities as in classical Be stars 
are found in a few cases of \be\ stars, but not much information is available 
or else is controversial, as in the case of HD\,45677. A common characteristic 
of \be\ and classical Be stars is certainly the non-sphericity of their 
circumstellar envelopes. Both groups appear to possess disk-like envelopes 
which for \be\ stars, but not for classical Be stars,
are dense enough to allow dust formation.

\acknowledgements{I would like to thank the organizers of the colloquium for
generously granting travel funds. I also would like to thank B. Wolf for 
making the spectrum of R\,50 available to me.
}

\section*{Discussion}
\noindent
{\bf Harmanec:} 1. What is known about the variability of B[e] stars? 2. How
did you derive the luminosities of B[e] stars used to plot particular stars into
the HR diagram? Cannot these luminosities and also $v \sin i$ values refer to
optically thick inner parts of their envelopes (i.e. pseudophotospheres) rather
than genuine photospheres?

\noindent
{\bf Zickgraf:} 1. For several galactic B[e] stars variability has been
observed. The B[e] supergiants in the Magellanic Clouds, however, in general do
not show pronounced variations. An exception is R~4 which is a kind of
B[e]/LBV-type star.
2. This question should be answered by J. Zorec who derived the stellar
parameters.

\noindent
{\bf Zorec:} The key parameter for luminosity estimation is the stellar 
distance.
The method used to derive the distance of some stars with the  B[e] phenomenon
is published in the proceedings of the ``B[e] stars'' workshop  (eds. A.M. 
Hubert and C. Jaschek). The basic assumptions I made are: 
(a) The spectrum of the star
underneath the circumstellar gas and dust envelope, determined using either the
BCD spectrophotometry or excitation arguments for the observed intensity of
visible emission lines, is assumed to correspond to a ``normal-like'' stellar
energy distribution.
(b) There is an emission component in the visible continuum spectrum 
resembling 
the one observed in classical Be stars. The amount of this emission, and the
intrinsic reddening that is associated with it, are estimated from the second
component of the Balmer discontinuity, as in classical Be stars.
(c) I took into account the UV-visible dust absorption due to the circumstellar
dust envelope and the corresponding re-emission of energy in the far-IR.
(d) The interstellar absorption as a function of the distance in the direction
of the star was determined as carefully as possible. Then, from an iterative
procedure to get a description of the amount of energy absorbed and re-emitted
by the gas-dust circumstellar envelope, which is also 
consistent with the excitation
produced by the underlying object, it is possible to obtain an estimate of the
circumstellar dust $E(B-V)$ component as well as the interstellar $E(B-V)$ and
consequently the stellar distance. 

The energy integrated over the whole spectral range is a quantity which is
treated in each iteration step. So, when you stop the iteration you also get the
right apparent bolometric luminosity which can be straightforwardly transformed
into an absolute bolometric luminosity. The method produced distances which
are quite comparable with the distances obtained from HIPPARCOS parallaxes, 
when they existed.

\noindent
{\bf Najarro:} In your color-color viewgraph there seems to be a clear separation
between emission line stars and B[e]. I was wondering if some of the B[e]s on
the lower edge of your diagram could just be LBV like stars (e.g. HD316285 and
R4) with such dense winds that the bound-free and free-free emission from the 
wind
could simulate the B[e] IR excess. A good way to test it would be to overplot
the color-color values for W-R stars. Have you tested that?

\noindent
{\bf Zickgraf:} No, I have not compared the IR colours of B[e] supergiants with
those of W-R stars. However, it seems difficult to produce such excesses with
(f-f)-(f-b) emission as observed in the B[e] region of the $(J-H)-(H-K)$ 
diagram.

\end{document}